\documentclass[12pt]{article}

\usepackage{newtxtext,newtxmath}

\usepackage{graphicx}

\usepackage[letterpaper,margin=1in]{geometry}

\renewenvironment{abstract}
	{\quotation}
	{\endquotation}

\date{}

\makeatletter
\renewcommand{\fnum@figure}{\textbf{Figure \thefigure}}
\renewcommand{\fnum@table}{\textbf{Table \thetable}}
\makeatother

\usepackage{scicite}

\usepackage{url}

\def\scititle{
	Nonlocal photonic time crystals: Infinite momentum bandgaps with minimal modulation speed and strength
}
\title{\bfseries \boldmath \scititle}

\author{
	Mohammadreza~Salehi$^{1\dagger}$,
	Matteo~Ciabattoni$^{1\dagger}$,
	Francesco~Monticone$^{1\ast}$\and
	\small$^{1}$School of Electrical and Computer Engineering, Cornell University, Ithaca, NY 14853, USA.\and
	\small$^\ast$Corresponding author. Email: francesco.monticone@cornell.edu\and
	\small$^\dagger$These authors contributed equally to this work.
}

\begin{document} 

\maketitle

\begin{abstract} \bfseries \boldmath
For over a decade, photonic time crystals have promised access to novel and exotic optical phenomena, offering fundamentally new ways to manipulate classical and quantum light. Central to these capabilities is the emergence of momentum bandgaps---the counterpart of the more familiar frequency bandgaps in spatial crystals---which have proven difficult to observe experimentally due to the combined need for high modulation speed and strength. To date, these requirements have all but hindered the development of time crystals at optical frequencies. Here, we show that the stringent modulation-speed requirement is a direct consequence of the Manley–Rowe relations governing conventional modulation schemes. We further demonstrate that modulating the plasma frequency of a Lorentz-dispersive material overcomes this limitation. Incorporating a specific form of spatial nonlocality (spatial dispersion) into this already temporally nonlocal (frequency dispersive) framework removes all remaining constraints, enabling momentum bandgaps of infinite extent---in both frequency and momentum---with arbitrarily small modulation speeds and strengths.
\end{abstract}

\noindent
There are not many physical phenomena at the frontier of modern research that can be understood by revisiting a childhood memory, yet parametric resonance is one of them. Arising from a periodic temporal modulation of a system parameter, the phenomenon underlies the familiar trick by which children make playground swings go higher. When a child rocks back and forth at appropriate points in the swing trajectory (twice per cycle), they periodically raise and lower the pendulum's center of mass. This ``pumping'' of the swing \cite{Ref1, Ref2} at twice its natural frequency produces an exponential amplification of the oscillation amplitude, in contrast to the linear growth associated with a regular resonance, in which an external force drives the swing at its natural frequency. The same principle has also been used for centuries to amplify the oscillations of the giant censer suspended in the transept of Spain's cathedral of Santiago de Compostela \cite{Ref3}. And today, after decades of research exploiting parametric effects in optical and microwave amplifiers \cite{boyd2020nonlinear,Ref20}, parametric resonance is attracting renewed attention in the field of electromagnetics and optics through the concept of photonic time crystals (PTCs) \cite{Ref4, Ref5, Ref5-1, Ref6}. Inspired in part by earlier work on time crystals in other areas of physics \cite{Ref37,Ref41}, PTCs are so named because their optical properties (such as the permittivity) are periodically modulated in time. This temporal periodicity gives rise to unusual wave phenomena, such as momentum bandgaps \cite{Ref4, Ref5-1}, amplification of spontaneous emission \cite{Ref6}, subluminal Cherenkov radiation \cite{Ref38}, superluminal momentum-gap solitons \cite{Ref39}, and an increasingly broad range of potential applications, including new forms of parametric amplification \cite{Ref6-1, Ref6-2, Ref6-3, Ref7, Ref8, Ref9}, broadband absorption \cite{Ref10, Ref11, Ref12}, and frequency conversion \cite{Ref13, Ref14, Ref15}.

To understand the challenges associated with realizing PTCs, a brief description of their electromagnetic response can offer relevant insight. A homogeneous PTC with a periodically modulated dielectric permittivity $\varepsilon(t)$ is governed by the time-varying wave equation $\partial^2\mathbf{D}/\partial t^2 - \nabla^2\mathbf{D}/\mu\varepsilon(t) = 0$, where $\mathbf{D}$ denotes the electric displacement field. For a transverse plane wave of arbitrary wavevector (linear momentum) $\mathbf{k}$, the spatial dependence reduces as $\nabla \rightarrow -i\mathbf{k}$, yielding the purely temporal equation $\partial^2\mathbf{D}/\partial t^2 + k^2\mathbf{D}/\mu\varepsilon(t) = 0$, where $k=|\mathbf{k}|$ is the wavenumber. This second-order ordinary differential equation with a time-varying coefficient is mathematically identical to that of a parametrically driven harmonic oscillator \cite{Ref40}; therefore, much like a periodically pumped swing, a PTC can exhibit parametric resonance. Consider a weak periodic modulation of the permittivity at frequency $\Omega$ around an average value $\varepsilon_\text{avg}$. In the absence of modulation, the medium supports two natural frequencies: $\omega^+(k)=k/\sqrt{\mu\varepsilon_\text{avg}}$ and $\omega^- (k)=-k/\sqrt{\mu\varepsilon_\text{avg}}$ (positive- and negative-frequency solutions). The temporal modulation perturbs and couples these natural modes/frequencies, with resonant coupling occurring when the modulation rate $\Omega$ bridges the separation between them, i.e., $\Omega \approx \omega^+(k)-\omega^-(k)=2\omega^+(k)$. Under this condition, the two modes hybridize into a pair of supermodes with complex eigenfrequencies, leading to exponentially growing and decaying amplitudes in time. The corresponding intervals of wavenumber $k$ that satisfy such a parametric resonance condition are referred to as momentum bandgaps. As attractive as these features may be, they come with significant challenges and practical limitations. First, momentum bandgaps large enough to be readily observed and exploited typically require order-unity modulation of the refractive index. This is because the width of any bandgap---whether in frequency or momentum space---is governed by the coupling coefficient $\kappa$, which scales with the strength of the temporal or spatial modulation. This requirement can be partially relaxed through dispersion engineering, for example, by creating flat bands as in \cite{Ref29}, but only to a limited extent \cite{Ref35}. Second, and more importantly, the resonance condition above dictates that momentum bandgaps occur only for waves with frequency $\omega = \Omega/2$ \cite{Ref8, Ref19}. While each of these issues can be addressed individually (e.g., in nonlinear optics, ultra-fast, albeit weak, modulations are routinely implemented \cite{boyd2020nonlinear}), realizing a wide momentum bandgap requires simultaneously satisfying both conditions: order-unity modulation strength and modulation rates twice the signal frequency. At optical frequencies, this corresponds to ultra-strong modulation at rates of hundreds of terahertz, which is a formidable challenge, well beyond the reach of existing technology \cite{Ref5-1, Ref18}.

Here, we address both of these fundamental obstacles that have so far hindered the realization of PTCs, potentially changing the paradigm by which these systems, and other classes of time-varying systems, are implemented. First, we introduce the concept of active pumping (in contrast to conventional ``reactive'' pumping) and show that it gives rise to a fundamentally new form of parametric resonance that does not require a modulation frequency $\Omega$ greater than the signal frequency. We identify materials and systems with a specific form of temporal dispersion as a suitable platform for active modulation and elucidate the underlying physics. Remarkably, combining this concept with suitable spatial dispersion (nonlocality) yields a recipe for what may be viewed as the ultimate momentum bandgap: one that is, in principle, \emph{infinitely wide} in both frequency and momentum, while requiring \emph{arbitrarily small} modulation frequencies and strengths. We demonstrate these concepts through theory, simulations, and proof-of-concept experiments.

\subsection*{Reactive vs Active Pumping}
The first step toward the ambitious goal of realizing the ultimate momentum bandgap is to determine whether, and under what conditions, a parametric resonance can be achieved with arbitrarily slow modulation. To address this question, we begin with a discussion of lumped-element parametric circuit resonators, drawing on the extensive literature on microwave circuit amplifiers, before returning to the framework of waves in media. The lumped-element equivalent of the time-varying wave equation introduced above is the equation governing the charge $q$ on a time-varying capacitor in an LC circuit: $d^2 q/dt^2 + q/LC(t) = 0$. As before, parametric resonance occurs when the modulation couples the natural modes of the circuit, which in this case are again the positive- and negative-frequency solutions associated with the same natural oscillation: $\Omega=\omega^+ - \omega^-=2\omega^+$. Accordingly, a conventional, or ``degenerate,'' parametric resonator \cite{Ref20} tuned to a signal frequency $\omega_\text{s}=\omega^+$ amplifies that signal when a reactive element (usually the capacitance) is modulated at twice the signal frequency, $\Omega=2\omega_\text{s}$. 

A ``non-degenerate'' parametric resonator \cite{Ref20, Ref21} relaxes this modulation speed requirement to an extent. This is a more general form of parametric resonator, with at least two resonant circuits coupled by a modulated reactive element, as shown in Fig.~\ref{fig1}A. Due to its multi-resonant nature, it supports parametric interactions between the natural mode tuned at the signal frequency $\omega_\text{s}$ and another natural mode at an ``idler'' frequency $\omega_\text{i}$. Floquet analysis shows that parametric resonance can now occur between the positive and negative frequencies belonging to different natural oscillations. The parametric resonance condition becomes: $\Omega=\omega_\text{s}^+-\omega_\text{i}^-=\omega_\text{s}+\omega_\text{i}$ (supplementary text S1). While the modulation speed $\Omega$ is still necessarily greater than the signal frequency $\omega_\text{s}$, it no longer needs to equal $2\omega_\text{s}$ since $\omega_\text{i}$ can be chosen independently. 

Figure~\ref{fig1}B illustrates the various parametric interactions supported by the non-degenerate parametric resonator of Fig.~\ref{fig1}A. Evidently, not all such interactions are of resonant nature. The \textit{counter-oscillating} interactions (between positive and negative frequencies) described thus far fall under the resonant category. By contrast, \textit{co-oscillating} interactions---where the modulation couples the \emph{positive} frequencies of different modes, $\Omega=\omega_\text{s}^+-\omega_\text{i}^+=\omega_\text{s}-\omega_\text{i}$---are non-resonant. As further discussed below, such interactions enable frequency conversion but not \emph{exponential} amplification, and therefore, when translated to a propagating-wave scenario, cannot open momentum bandgaps. Unfortunately, the parametric interaction best suited to our goal is precisely of this co-oscillating type, since the condition $\Omega = \omega_\text{s} - \omega_\text{i}$ would permit an arbitrarily small modulation frequency when $\omega_\text{s}$ and $\omega_\text{i}$ are close. This raises a central question: Can one devise a new, more sophisticated circuit configuration incorporating time-varying capacitors in which such co-oscillating interactions become resonant?  

The answer to this question is given by the Manley-Rowe relations \cite{Ref22, Ref20}: for time-varying systems based on the modulation of capacitors, or indeed any reactive elements, the answer is a resounding no. The Manley-Rowe relations are a set of fundamental power conservation laws, of great relevance for both microwave engineering and nonlinear optics \cite{boyd2020nonlinear,Ref20}, that apply to generic parametric devices consisting of nonlinear reactances. While our discussion so far focused on linear systems, this is relevant because time-varying capacitances in parametric amplifiers are typically implemented using nonlinear capacitors. This is because, under standard small-signal assumptions, a nonlinear capacitor follows a linearized time-varying equation \cite{Ref20}. The linearized model remains valid as long as the signal amplitude is much smaller than that of the pump. Under parametric resonance, however, the signal grows exponentially and eventually invalidates the small-signal assumption. The response then becomes strongly nonlinear, the system reaches a steady state, and the capacitor acts as a frequency mixer, exchanging power among different frequencies and harmonics. When the signal (at frequency $\omega_\text{s}$) and the pump (at frequency $\Omega$) undergo nonlinear mixing, new spectral components are generated at frequencies $n\omega_\text{s}+m\Omega$ ($n,m \in \mathbb{Z}$). Let $P_{nm}$ denote the power delivered to the nonlinear capacitor at the frequency $n\omega_\text{s}+m\Omega$. Because the capacitor is purely reactive and cannot dissipate or generate net power, overall power conservation requires $\sum_{n=-\infty}^{+\infty}\sum_{m=-\infty}^{+\infty}P_{nm}=0$. The Manley-Rowe relations follow from this constraint. By invoking the characteristic equations of a capacitor (or inductor) and performing simple algebraic manipulations, one obtains \cite{Ref22, Ref20}:
\begin{subequations} \label{eq1}
\begin{equation} \label{eq1a}
\sum_{n=0}^{+\infty}\sum_{m=-\infty}^{+\infty} \frac{n P_{nm}}{n\omega_\text{s}+m\Omega}=0,
\end{equation}
\begin{equation} \label{eq1b}
\sum_{m=0}^{+\infty}\sum_{n=-\infty}^{+\infty} \frac{m P_{nm}}{n\omega_\text{s}+m\Omega}=0.
\end{equation}
\end{subequations}
\noindent As commonly done in the application of the Manley-Rowe relations to microwave or optical amplifiers, we assume that three frequencies participate in the nonlinear mixing, namely $\omega_\text{s}$, $\omega_\text{i}$, and $\Omega$. Although the time-varying circuit in Fig.~\ref{fig1}A does not explicitly include the pump, the modulation is, in practice, implemented through additional circuitry that couples the pump to the physical nonlinear element, thereby introducing $\Omega$ as a third spectral component. When the pump is operated at $\Omega=\omega_\text{s}+\omega_\text{i}$, the only non-zero terms in Eq. \ref{eq1}  are those with $n=\pm1, m=0$ and $n=0, m=\pm1$ and $n=-m=\pm1$. Then, considering that $P_{1,0}=P_{-1,0}=P_\text{s}$ and $P_{1,-1}=P_{-1,1}=P_\text{i}$ and $P_{0,1}=P_{0,-1}=P_\text{p}$, we obtain
\begin{subequations} \label{eq2}
\begin{equation} \label{eq2a}
\frac{P_\text{s}}{\omega_\text{s}}-\frac{P_\text{i}}{\omega_\text{i}}=0,
\end{equation}
\begin{equation} \label{eq2b}
\frac{P_\text{p}}{\Omega}+\frac{P_\text{i}}{\omega_\text{i}}=0.
\end{equation}
\end{subequations}
\noindent Since the pump supplies power to the nonlinear element, $P_\text{p}>0$. Equation~\ref{eq2b} therefore implies $P_\text{i}<0$, and Eq.~\ref{eq2a} then requires $P_\text{s}<0$. Thus, the nonlinear capacitor extracts power from the pump and delivers it to both the signal and idler. This behavior is consistent with parametric resonance, in which natural modes exhibit exponential growth without requiring sources at $\omega_s$ or $\omega_i$ (sources are irrelevant to the natural frequencies of a system). Being a steady-state framework, the Manley–Rowe relations can only be applied after this exponential growth inevitably saturates, at which point they correctly predict that the self-sustained modes extract power from the pump, which is the sole power-generating component. In contrast, when applied to co-oscillating interactions, the resulting Manley–Rowe relations are incompatible with the dynamics of parametric resonance. For co-oscillating interactions ($\Omega=\omega_\text{s}-\omega_\text{i}$), the same terms in Eq.~\ref{eq1} are non-zero as in the previous case. The crucial difference, however, is that the index set $n=-m=1$ now corresponds to $\omega_\text{i}$ rather than $-\omega_\text{i}$. Using the same power equalities as before, the Manley-Rowe relations reduce to
\begin{subequations} \label{eq3}
\begin{equation} \label{eq3a}
\frac{P_\text{s}}{\omega_\text{s}}+\frac{P_\text{i}}{\omega_\text{i}}=0,
\end{equation}
\begin{equation} \label{eq3b}
\frac{P_\text{p}}{\Omega}-\frac{P_\text{i}}{\omega_\text{i}}=0.
\end{equation}
\end{subequations}
\noindent $P_\text{p}$ is again positive, which directly implies $P_\text{s}<0$, but $P_\text{i}>0$. This means that the nonlinear capacitor extracts power from the idler $\omega_\text{i}$ (and the pump) and delivers it to signal $\omega_\text{s}$. Again, parametric resonance entails the emergence of exponentially growing modes sustained solely by the pump, without any additional sources. However, Eq.~\ref{eq3} necessarily implies the presence of an additional power-generating contribution at $\omega_\text{i}$ (since $P_\text{i}>0$). This inconsistency demonstrates that, for a modulated reactive element, the Manley–Rowe relations render co-oscillating interactions and parametric resonance mutually exclusive. Thus, rather than producing parametric resonance, co-oscillating interactions between positive frequencies lead to frequency conversion (which may be accompanied by some amplification, \cite{Ref20}). In propagating-wave systems, this effect is certainly useful, and, when applied in the form of a traveling-wave modulation, can also break reciprocity \cite{Ref16, Ref17}; however, because it does not produce exponential growth, it cannot open a momentum bandgap.

\subsubsection*{Active Pumping: The Role of Dispersion}
All of the above discussion hinges on the assumption that a reactive element is being modulated. A \textit{reactive pumping scheme}, whether degenerate or non-degenerate, is therefore fundamentally incapable of producing co-oscillating parametric resonance and, hence, of opening momentum bandgaps with small modulation frequencies. What, then, would an appropriate pumping scheme look like? One possibility is to consider more elaborate networks of reactive elements. However, we demonstrate (supplementary text S3) that the Manley-Rowe relations apply, in a similar fashion, to any system of mutually coupled nonlinear capacitors or inductors, making this option ineffective as well . Moreover, modulating a resistor is not a viable alternative either, as it cannot bring about amplification \cite{Ref11}. The remaining possibility is to modulate a fundamentally different kind of circuit element, namely, a dependent voltage or current source. This \textit{active pumping scheme} can be shown to break the Manley-Rowe relations (supplementary text S4), removing the fundamental roadblock to co-oscillating parametric resonance. This result is already significant in its own right, as it represents a fundamentally new class of parametric circuit amplifier, beyond the degenerate and nondegenerate amplifiers that have been well established in the literature for decades. Here, however, our main interest lies in understanding what this concept of active pumping implies in the broader context of time-varying material responses and wave systems.

\begin{figure} 
	\centering
	\includegraphics[width=1\textwidth]{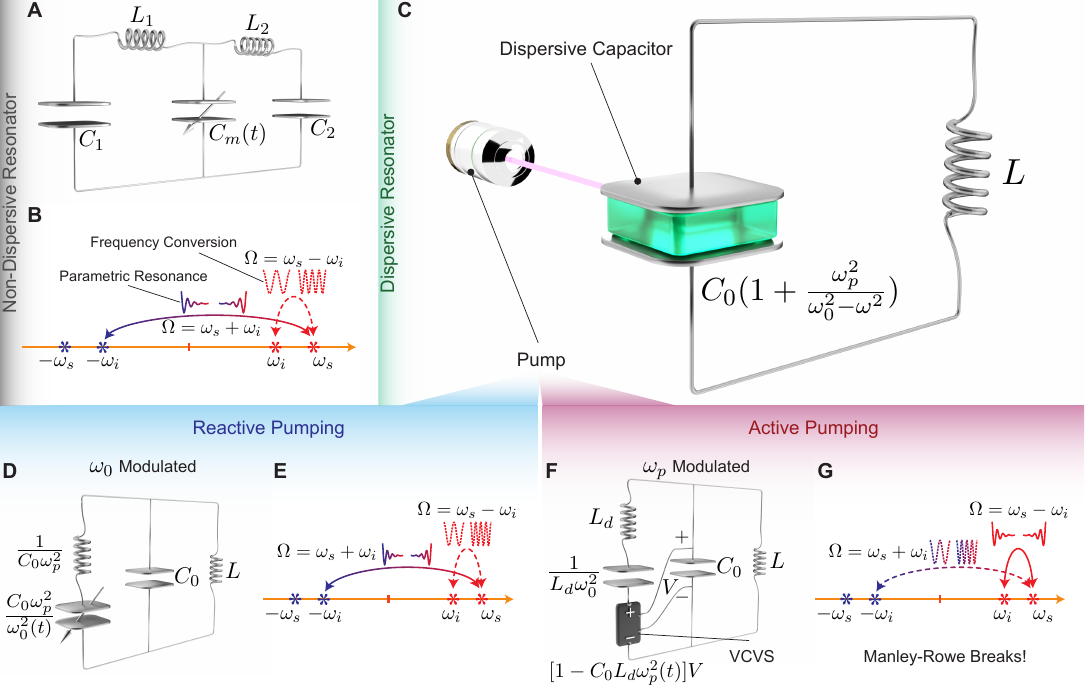} 

	\caption{\textbf{Parametric interactions in non-dispersive and dispersive multi-resonant circuits.}
		(\textbf{A}) A multi-resonant circuit featuring a time-modulated, non-dispersive varactor $C_\text{m}(t)$ with modulation frequency $\Omega$. (\textbf{B}) Diagram of the parametric interactions for the resonator in (A). The natural frequencies of the unmodulated circuit (i.e., if $C_\text{m}(t)$ equals its average value) are denoted $\omega_\text{s}, \omega_\text{i}$. The counter-oscillating interaction ($\Omega = \omega_\text{s}+\omega_\text{i}$) gives rise to parametric resonance, while the co-oscillating interaction ($\Omega = \omega_\text{s}-\omega_\text{i}$) results in frequency conversion, consistent with the Manley-Rowe relations. (\textbf{C}) A dispersive LC resonator featuring a Drude-Lorentz-type capacitance, $C_0[1+\omega_\text{p}^2/(\omega_0^2-\omega^2)]$. Owing to the material resonance, this is also a multi-resonant circuit, as it combines an ordinary LC resonator with a Lorentz harmonic oscillator representing the polarization dynamics of the dielectric. (\textbf{D}) Circuit equivalent of (C) when $\omega_0$ is modulated (reactive pumping). (\textbf{E}) Diagram of the parametric interactions for the resonator in (D). (\textbf{F}) Circuit equivalent of (C) when $\omega_\text{p}$ is modulated (active pumping), featuring a voltage-controlled voltage source (VCVS). The inductance $L_\text{d}$ is arbitrary. (\textbf{G}) Diagram of the parametric interactions for the resonator in (F). In contrast to the reactive case, the co-oscillating interaction is now resonant.}
	\label{fig1} 
\end{figure}

To this end, we first consider an LC resonator with a dispersive capacitance of Drude–Lorentz form $C(\omega)=C_0[1+\omega_\text{p}^2/(\omega_0^2-\omega^2)]$, where $C_0$ is a constant and $\omega_0, \omega_\text{p}$ denote the resonant and plasma frequencies of the dispersive material, respectively (in a Lorentz medium, $\omega_\text{p}$ is proportional to the oscillator strength of the resonance). Because of this internal material resonance, the circuit (Fig.~\ref{fig1}C) supports two distinct natural modes (at $\omega_\text{s}$ and $\omega_\text{i}$), as it is mathematically equivalent to an ordinary LC resonator coupled with a Lorentz harmonic oscillator representing the polarization dynamics of the dielectric. To apply the concepts of active and reactive pumping, we rigorously show (supplementary text S2) that this dispersive LC resonator is equivalent to the circuit networks of Figs.~\ref{fig1}D and \ref{fig1}F, depending on which material parameter is modulated. Specifically, modulating the resonant frequency $\omega_0$ is equivalent to modulating an ordinary non-dispersive capacitor, as shown in Fig.~\ref{fig1}D, and therefore constitutes a form of reactive pumping. By contrast, modulating the plasma frequency $\omega_\text{p}$ is equivalent to modulating a voltage-controlled voltage source (VCVS), as illustrated in Fig.~\ref{fig1}F, and therefore represents the material analogue of active pumping. 

Floquet analysis confirms that these two cases lead to fundamentally different parametric interactions. Modulating $\omega_0$ produces the conventional counter-oscillating resonance discussed above. In contrast, modulating $\omega_\text{p}$ results in co-oscillating parametric resonance and, hence, exponential amplification even when the modulation frequency is $\Omega=\omega_\text{s}-\omega_\text{i}$ (supplementary text S2). The diagrams of parametric interactions for the cases of time-modulated $\omega_0$ and $\omega_\text{p}$ are shown in Figs.~\ref{fig1}E and \ref{fig1}G, respectively. For a discussion of techniques used for modulating $\omega_0$ and $\omega_\text{p}$ in dispersive materials, we refer the reader to \cite{Ref5-1, Ref5-2, Ref5-3, Ref5-4, Ref5-5, Ref5-6, Ref5-7}.

\subsection*{Dispersion and Nonlocality: Infinite Momentum Bandgaps}
Having identified a specific form of time-varying frequency dispersion as the essential ingredient for realizing co-oscillating parametric resonances, we turn our attention to its implications for momentum bandgaps in PTCs. Consider an $x$-polarized wave propagating along the $z$ direction in a lossless, isotropic, dispersive material of Drude-Lorentz type, $\varepsilon(\omega)=1+\omega_\text{p}^2/(\omega_0^2-\omega^2)$. Under active pumping, that is, when the plasma frequency is modulated, the wave satisfies the following system of second-order differential equations (supplementary text S5):
\begin{subequations} \label{eq4}
\begin{equation} \label{eq4a}
\frac{\partial^2E_x}{\partial t^2} + k_z^2 c_0^2 E_x = -\frac{1}{\varepsilon_0}\frac{\partial^2P_x}{\partial t^2},
\end{equation}
\begin{equation} \label{eq4b}
\frac{\partial^2P_x}{\partial t^2} + \omega_0^2 P_x = \varepsilon_0\omega_\text{p}^2(t) E_x,
\end{equation}
\end{subequations}
\noindent where $P_x$ is the $x$ component of the polarization density vector, $k_z$ is the wavenumber in $z$ direction, and $c_0$ is the speed of light in vacuum. Being a fourth-order equation, the dispersion relation ($\omega$ vs $k_z$) of the modes features two pairs of positive- and negative-frequency branches. In drastic contrast to conventional, reactively pumped PTCs, active pumping resonantly couples \emph{positive} branches whenever their frequency separation equals $\Omega$. Since these branches can approach one another arbitrarily closely, the required modulation frequency can, in principle, be arbitrarily small. This results in a new type of momentum bandgap, formed by the resonant coupling between co-propagating states. Such \textit{co-propagating} momentum bandgaps have recently been observed numerically \cite{Ref12, Ref23, Ref24} and can now be fully explained by our theoretical framework as a consequence of active pumping that breaks the constraints imposed by the Manley-Rowe relations.  

\begin{figure} 
	\centering
	\includegraphics[width=1\textwidth]{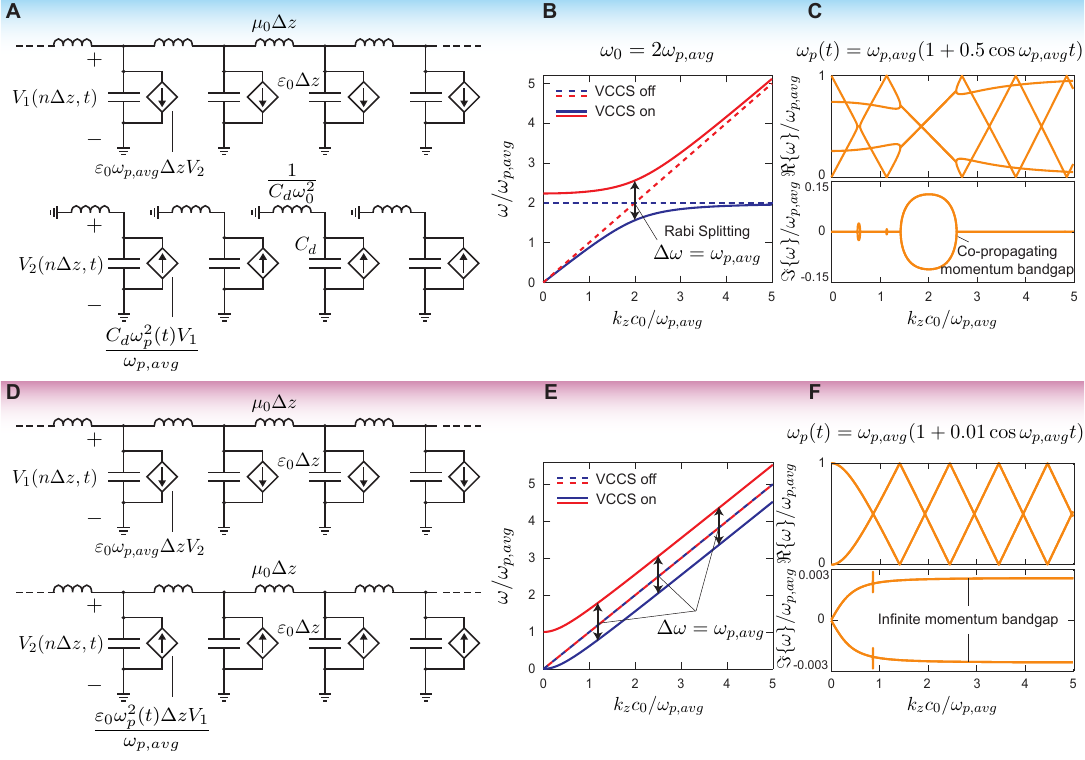} 

	\caption{\textbf{Co-propagating momentum bandgaps in dispersive and nonlocal systems.}
		(\textbf{A}) Exact transmission-line equivalent of a frequency-dispersive material of Drude-Lorentz type described by Eq. \ref{eq4}. Light–matter interaction is modeled through the coupling between a transmission line and a series of localized resonators, mediated by voltage-controlled current sources (VCCSs). Circuit element values for a single unit cell of length $\Delta z$ are specified. $\omega_\text{p,avg}$ denotes the average value of $\omega_\text{p}(t)$, and the capacitance $C_\text{d}$ is arbitrary. (\textbf{B}) Band diagram of the unmodulated structure in (A) for $\omega_\text{p}(t) = \omega_\text{p,avg}$ and the specified $\omega_0$. The dashed lines show the uncoupled photonic and matter-based states in the absence of light-matter coupling (VCCSs off), while the solid lines show the hybridized states arising from the coupling. (\textbf{C}) Band diagram of the time-modulated structure for the specified $\omega_\text{p}(t)$, showing a finite co-propagating momentum bandgap. (\textbf{D}) Proposed frequency- and spatially dispersive structure, consisting of two identical transmission lines coupled through the same mechanism, designed to produce parallel dispersion bands. (\textbf{E}) Band diagram of the unmodulated structure in (D) for $\omega_\text{p}(t) = \omega_\text{p,avg}$. (\textbf{F}) Band diagram of the time-modulated structure for the specified $\omega_\text{p}(t)$, showing an infinite co-propagating momentum bandgap.}
	\label{fig2} 
\end{figure}

However, co-propagating momentum bandgaps in dispersive PTCs are by no means wide. To understand why, it is useful to consider the transmission line equivalent of Eq. \ref{eq4}. A simple transformation of variables, $E_x=V_1$, ${\partial P_x}/{\partial t}=\varepsilon_0\omega_\text{p,avg}V_2$, where $\omega_\text{p,avg}$ is the average value of $\omega_\text{p}(t)$, transforms Eq.~\ref{eq4} into the equations governing the equivalent network shown in Fig.~\ref{fig2}A (supplementary text S5): a transmission line modeling electromagnetic wave propagation coupled to localized resonators representing the Drude-Lorentz material resonance. The variables $V_1$ and $V_2$ correspond, respectively, to the voltage on the transmission line and to the voltage across the capacitors of the localized resonators. These two subsystems are coupled through voltage-controlled current sources (VCCSs), with those in the lower branch being time-modulated. We emphasize that this network model is exact: the original equations map directly onto the equations of the network, with no approximations. We also note that although VCVSs were employed in the previous section to illustrate the concept of active pumping, VCCSs are equally suitable, as they also break the Manley-Rowe relations (supplementary text S4). 

Figure~\ref{fig2}B shows the dispersion diagram of the unmodulated structure. In the absence of the VCCSs, the transmission line and the localized resonators would correspond, respectively, to the uncoupled photonic and matter-excitation states (e.g., a photon and a plasmon), whose dispersions are indicated by the oblique and horizontal dashed lines. The VCCSs represent light-matter coupling, which gives rise to hybridized polaritonic states shown by the solid lines. The two branches exhibit an avoided crossing, or Rabi splitting \cite{Ref25}. This hybridization also explains why the resulting co-propagating momentum bandgaps remain limited in size. Because the separation between the two polaritonic branches varies strongly with frequency and wavevector, no single modulation frequency can resonantly couple them over a broad region of the dispersion diagram. This is evident from the dispersion diagram (Fig.~\ref{fig2}C) of the time-varying structure, where the modulation frequency was chosen to match the band separation at the location of the avoided crossing. A co-propagating momentum bandgap appears, but only in a relatively narrow neighborhood of that point.

The line of argument above presents us with a surprisingly simple and physically intuitive strategy for achieving arbitrarily large momentum bandgaps. The limitation of the structure in Fig.~\ref{fig2}A is that it embodies a hybridization process between fundamentally different types of states, one propagative (photonic) and the other localized (matter-based). Consequently, the two resulting dispersion bands cannot remain uniformly separated. A natural improvement on this configuration would therefore be a structure that allows for hybridization between two identical propagating states. In this case, the coupling-induced splitting would be uniform across all values of $\omega$ and $k_z$, yielding two parallel dispersion bands. Figure~\ref{fig2}D shows such a structure, in which two identical transmission lines are coupled through VCCSs, with those in the lower branch again being time-modulated (note that the use of VCCS coupling is essential: if the two transmission lines were instead coupled through conventional capacitive or inductive coupling, the resulting dispersion would be drastically different). As is readily apparent, this configuration corresponds to the original structure in Fig.~\ref{fig2}A in which the individual resonators are connected to form a second wave-carrying transmission line, with its parameters tuned to match the dispersion of the first. 

To move back to a wave-in-material scenario, we use the same transformation of variables as before, $E_x=V_1$, ${\partial P_x}/{\partial t}=\varepsilon_0\omega_\text{p,avg}V_2$, arriving at the material counterpart of this configuration (supplementary text S5):
\begin{subequations} \label{eq5}
\begin{equation} \label{eq5a}
\frac{\partial^2E_x}{\partial t^2} + k_z^2 c_0^2 E_x = -\frac{1}{\varepsilon_0}\frac{\partial^2P_x}{\partial t^2},
\end{equation}
\begin{equation} \label{eq5b}
\frac{\partial^2P_x}{\partial t^2} + k_z^2 c_0^2 P_x = \varepsilon_0\omega_\text{p}^2(t) E_x.
\end{equation}
\end{subequations}
\noindent 
Comparing these equations with Eq.~\ref{eq4}, we see that this new medium must exhibit not only frequency dispersion, but also spatial dispersion, i.e., nonlocality in both time and space, as implied by the wavevector-dependent term in the differential equation for the the polarization density, Eq.~\ref{eq5}b. Figures~\ref{fig2}E-F show the dispersion diagram for electromagnetic waves in such a medium or its exact transmission-line model. As expected, the modes of the unmodulated material ($\omega_\text{p}(t)=\omega_\text{p,avg}$) are indeed parallel, i.e., their frequencies are separated everywhere by exactly $\Delta \omega=\omega_\text{p,avg}$ (supplementary text S5) as shown by Fig.~\ref{fig2}E. Remarkably, modulating $\omega_\text{p}$ at a frequency $\Omega=\omega_\text{p,avg}$ creates a co-propagating momentum bandgap spanning the \emph{entire} range of $\omega$ and $k_z$ (Fig.~\ref{fig2}F). In other words, the material exponentially amplifies waves of \emph{arbitrary} frequency and momentum. Moreover, both the modulation frequency and modulation strength may be made arbitrarily small. This represents a dramatic improvement over previous work on PTCs, including recent results \cite{Ref29} that numerically demonstrated momentum bandgaps over a wider wavevector range, but occurring only at a single fixed frequency and still requiring a modulation frequency twice that of the propagating wave.

\begin{figure} 
	\centering
	\includegraphics[width=0.94\textwidth]{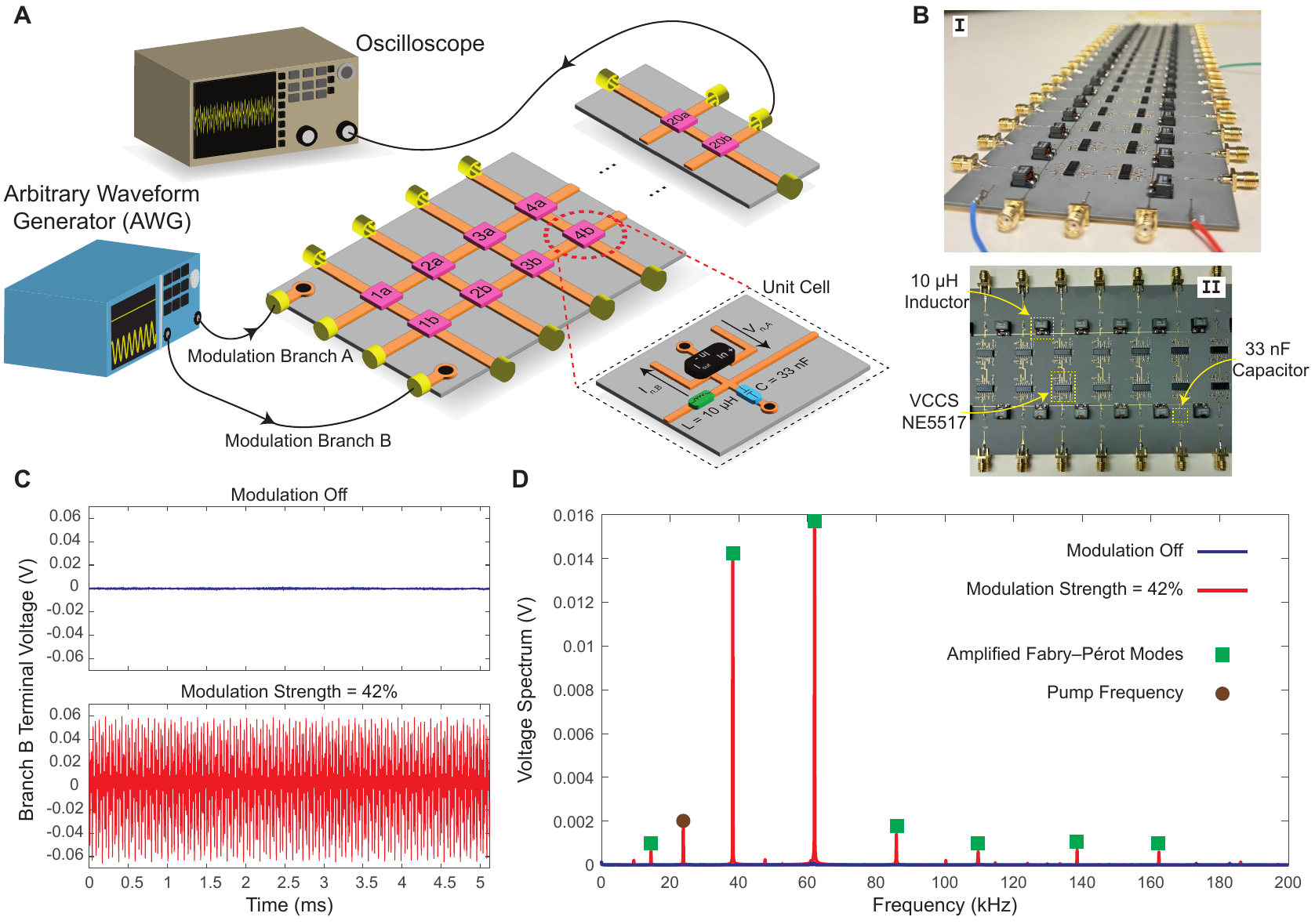} 

	\caption{\textbf{Experimental demonstration of an infinite momentum bandgap.}
		(\textbf{A}) Illustration and (\textbf{B}) photograph of the experimental setup, consisting of a Fabry–Pérot resonator composed of 20 unit cells of the structure in Fig.~\ref{fig2}D, with open-circuit terminations at all end ports. The VCCSs in branch A are sinusoidally modulated at $\Omega=23.8~\text{kHz}$ about an average value equal to that of the unmodulated VCCSs in branch B. When the modulation strength exceeds a threshold value (due to losses), all Fabry–Pérot modes of the structure experience exponential amplification until the VCCSs reach saturation. The voltage is then measured at one of the end ports in branch B using an oscilloscope. (\textbf{C}) $5~\text{ms}$-long snapshots of the measured voltage are shown for two cases: with the modulation turned off (top), and with the modulation strength above the amplification threshold (bottom). (\textbf{D}) Corresponding spectra of the measured signals in (C), showing strong ultra-broadband parametric amplification of the Fabry–Pérot modes, including at frequencies far exceeding $\Omega$. (The appearance of the pump frequency in the spectrum arises because the baseline DC offset in the VCCSs’ output is picked up and subsequently modulated by the VCCSs in the opposite branch.) In comparison, a conventional parametric amplifier or reactively pumped PTC would amplify only the Fabry–Pérot mode satisfying $\omega = \Omega/2$. }
	\label{fig4} 
\end{figure}

We validated these findings through a proof-of-concept implementation of the circuit in Fig.~\ref{fig2}D, using readily available low-frequency commercial VCCSs. The circuit comprises 20 unit cells and is terminated by open circuits at all ports, forming an artificial Fabry–Pérot resonator (Fig.~\ref{fig4}A) that supports a discrete set of wavenumbers and frequencies. In a conventional parametric amplifier or reactively pumped PTC, only the Fabry–Pérot mode whose frequency satisfies $\omega = \Omega/2$ is amplified. By contrast, the infinite momentum bandgap of our structure enables the amplification of all Fabry–Pérot modes, including those at frequencies far greater than the modulation frequency. The fabricated circuit shown in Fig.~\ref{fig4}B is modulated at $23.8~\text{kHz}$. Due to intrinsic losses in the discrete circuit elements---particularly the inductors---the modulation strength must exceed a threshold for exponential amplification to occur. As shown in movie S1, once this threshold is surpassed (corresponding to a modulation depth of $42\%$), all Fabry–Pérot modes undergo exponential growth until the VCCSs reach saturation, at which point the gain stabilizes. Figure~\ref{fig4}C presents a snapshot of the voltage measured at the end of the resonator after gain stabilization. The spectrum of the measured signal (Fig.~\ref{fig4}D) clearly demonstrates ultra-broadband amplification of the Fabry–Pérot modes, confirming the formation of an extremely wide momentum bandgap under low-frequency modulation.  

The nonuniform gain observed in this proof-of-concept experiment (unequal amplitudes of the peaks in Fig.~\ref{fig4}D) arises from variations in the values of the discrete circuit elements (e.g., inductors with a tolerance of $20\%$). These nonidealities perturb the dispersion relation, making the two branches no longer perfectly parallel, resulting in the observed gain nonuniformity. We also note that, at the very low frequencies considered here, the use of LC ladder networks is essential, as an equivalent physical transmission line would be impractically long. For future high-frequency implementations, however, the LC ladder network could be replaced by a physical waveguide, such as a microstrip line, which would provide much more uniform inductance and capacitance per unit length, thereby mitigating this issue. 

Finally, a natural question to ask is: what type of physical material or metamaterial is described by Eq.~\ref{eq5}? In the Fourier domain, Eq.~\ref{eq5} with constant $\omega_\text{p}$ corresponds to a medium with relative permittivity $\varepsilon(\omega, k_z)=1+\omega_\text{p}^2/(k_z^2 c_0^2-\omega^2)$. Applying the inverse Fourier transform yields the following spatio-temporal constitutive relation for the material (supplementary text S5):
\begin{equation} \label{eq6}
D_x(z,t) = \varepsilon_0 E_x(z,t)+\frac{\varepsilon_0 \omega_\text{p}^2}{2c_0}\int_{-\infty}^{+\infty}\int_{-\infty}^{+\infty} u(t-t'-\frac{|z-z'|}{c_0}) E_x(z',t') dt' dz',
\end{equation}
\noindent where $D_x(z,t)$ is the $x$ component of the electric displacement vector and $u(\cdot)$ is the Heaviside step function. At a given point in space and time, the material response is determined by the input electric field, not just at that point, but over a range of $(z,t)$ values; hence, the medium is indeed temporally and spatially nonlocal, with a peculiar type of space-time nonlocal kernel. Equation~\ref{eq6} further indicates that the material respects causality and special relativity (the nonlocal kernel is causal and does not transfer information faster than light in vacuum), suggesting that such a material may indeed be physically realizable. 

\begin{figure} 
	\centering
	\includegraphics[width=0.6\textwidth]{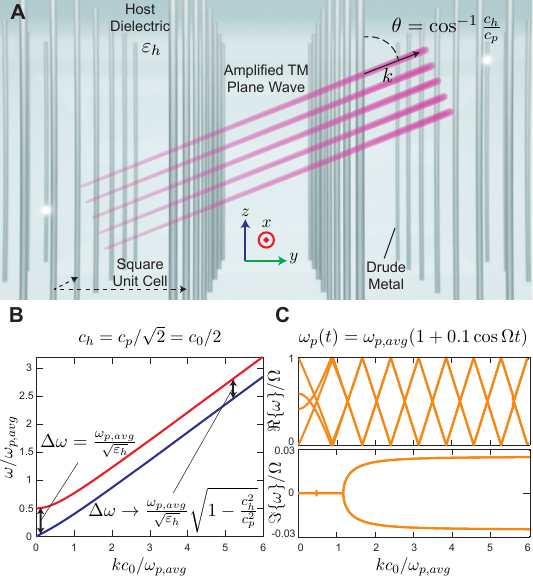} 

	\caption{\textbf{Infinite momentum bandgap in a nonlocal PTC metamaterial.}
		(\textbf{A}) Illustration of a metamaterial consisting of an array of long, thin metallic wires arranged in a square lattice and embedded in a host dielectric of relative permittivity $\varepsilon_\text{h}$. Under the long-wavelength approximation, the structure is characterized by an anisotropic effective permittivity with components $\varepsilon_{xx}=\varepsilon_{yy}=\varepsilon_\text{h}$ and $\varepsilon_{zz}(\omega, k_z)=\varepsilon_\text{h}+\omega_\text{p}^2/(k_z^2 c_\text{p}^2-\omega^2)$. (\textbf{B}) Band diagram for transverse-magnetic (TM) plane waves in the unmodulated metamaterial ($\omega_\text{p}(t)=\omega_{\text{p,avg}}$) propagating at an angle $\theta=\arccos(c_\text{h}/c_\text{p})$ with respect to the wires, where $c_\text{h}^2=1/(\mu_0 \varepsilon_0 \varepsilon_\text{h})$. The two bands are initially nonparallel but asymptotically approach a constant separation as $k\rightarrow\infty$. (\textbf{C}) Band diagram of the time-modulated structure for the specified $\omega_\text{p}(t)$, with $\Omega = \frac{\omega_\text{p,avg}}{\sqrt{\varepsilon_\text{h}}}\sqrt{1-c_\text{h}^2/c_\text{p}^2}$, showing a semi-infinite co-propagating momentum bandgap.}
	\label{fig3} 
\end{figure}

A promising candidate is a metamaterial consisting of parallel $z$-oriented thin metallic wires arranged in a square lattice in the $x$–$y$ plane, as illustrated in Fig.~\ref{fig3}A, which can be modeled, under the long-wavelength approximation, by an anisotropic effective permittivity with similar properties \cite{Ref27}. When the wires are embedded in a host dielectric of relative permittivity $\varepsilon_\text{h}$, the resulting effective permittivity tensor is diagonal, with components $\varepsilon_{xx}=\varepsilon_{yy}=\varepsilon_\text{h}$ and $\varepsilon_{zz}(\omega, k_z)=\varepsilon_\text{h}+\omega_\text{p}^2/(k_z^2 c_\text{p}^2-\omega^2)$ (supplementary text S6), with the $zz$-component having the same frequency- and spatially dispersive form as the desired medium. The constants $\omega_\text{p}$ and $c_\text{p}$ are determined by the structural parameters of the wire medium. For a time-modulated $\omega_\text{p}$, transverse magnetic plane waves (with field components $H_x$, $E_y$, and $E_z$) propagating in the $y-z$ plane at an angle $\theta$ with respect to the wires, satisfy the following system of equations (supplementary text S6): 
\begin{subequations} \label{eq7}
\begin{equation} \label{eq7a}
\frac{\partial^2 E_z}{\partial t^2}+k^2c_\text{h}^2E_z=-\frac{1}{\varepsilon_0\varepsilon_\text{h}}(\frac{\partial^2 P_{z,\text{wire}}}{\partial t^2}+k^2c_\text{h}^2 \cos^2 \theta P_{z,\text{wire}}),
\end{equation}
\begin{equation} \label{eq7b}
\frac{\partial^2 P_{z,\text{wire}}}{\partial t^2}+k^2c_\text{p}^2 \cos^2 \theta P_{z,\text{wire}}=\varepsilon_0\omega_\text{p}^2(t)E_z,
\end{equation}
\end{subequations}
\noindent where $c_\text{h}^2 = 1/(\mu_0 \varepsilon_0 \varepsilon_\text{h})$ and $P_{z,\text{wire}}$ denotes the contribution of the wires to the polarization density. In the time-invariant case (when $\omega_\text{p}(t) = \omega_\text{p,avg}$), achieving a dispersion diagram with parallel bands requires $k^2 c_\text{h}^2 = k^2 c_\text{p}^2 \cos^2 \theta$, which fixes the propagation angle to $\theta = \arccos (c_\text{h}/c_\text{p})$. However, because of the slightly different form of Eq.~\ref{eq7} compared to Eq.~\ref{eq5}, the resulting bands are not perfectly parallel over the entire spectrum, as shown in Fig.~\ref{fig3}B. Specifically, the band separation is initially larger, with $\Delta \omega = \omega_\text{p,avg}/\sqrt{\varepsilon_\text{h}}$ at $k = 0$, and gradually converges to the constant value $\Delta \omega = \frac{\omega_\text{p,avg}}{\sqrt{\varepsilon_\text{h}}}\sqrt{1-c_\text{h}^2/c_\text{p}^2}$ deeper within the band (supplementary text S6). Figure~\ref{fig3}C shows the dispersion diagram of the nonlocal PTC slowly modulated at a frequency $\Omega = \frac{\omega_\text{p,avg}}{\sqrt{\varepsilon_\text{h}}}\sqrt{1-c_\text{h}^2/c_\text{p}^2}$, demonstrating the emergence of a semi-infinite momentum bandgap in a metamaterial platform potentially within reach of current experimental capabilities at optical frequencies. 

\subsection*{Discussion and Outlook}

This work resolves the long-standing limitations associated with momentum bandgaps in PTCs, removing a major barrier to unlocking the full potential of time-varying photonics and electromagnetics. We have introduced the concept of active pumping to overcome the stringent modulation speed requirement of conventional PTCs, imposed by the Manley-Rowe relations, and identified a special class of frequency-dispersive media as a viable platform for its realization. Incorporating spatial dispersion (nonlocality in space) completes the picture and enables the realization of the ultimate momentum bandgap: infinitely wide in both frequency and momentum, and attainable with arbitrarily small modulation speed and strength. We have validated this first-of-its-kind nonlocal PTC both numerically at optical frequencies, demonstrating an infinite momentum bandgap in a time-varying wire-medium metamaterial, and experimentally through a low-frequency proof-of-concept based on time-modulated circuit networks. Looking ahead, we expect that nonlocal PTCs at RF and microwave frequencies could also be implemented using RFIC or MMIC technologies \cite{Ref28}, potentially operating at frequencies well beyond 100~GHz \cite{Ref36}. Such integrated implementations would ultimately realize the full potential of the new class of parametric circuit amplifier introduced here for communications and other applications.

From a broader scientific perspective, momentum bandgaps have long been heralded as a gateway to a wide range of novel and exciting optical phenomena. Nevertheless, the severity of their practical limitations has hindered their experimental realization at high frequencies. By overcoming these constraints, our work may help catalyze a new wave of research in time-varying photonics. Infinite momentum bandgaps open new possibilities for advanced classical and quantum light–matter interactions \cite{Ref32} and create new opportunities for wave amplification, spectral control, and high-resolution imaging \cite{Ref5, Ref10, Ref33, Ref34, Ref35}, both in optics and in other domains of wave physics.


\clearpage 

%
\bibliography{science_template} 
\bibliographystyle{sciencemag}

%
%
%
%
%
%


\section*{Acknowledgments}


\paragraph*{Author contributions:}
F.M. conceived and supervised the study. M.S. developed the theory, performed theoretical calculations and numerical simulations, designed the fabricated structure and the experimental setup, analyzed the experimental data, and led the writing of the manuscript with contributions from all authors. M.C. designed and assembled the fabricated structure, designed and built the experimental setup, performed numerical simulations and measurements, and analyzed the experimental data. All authors participated in the conceptual development, discussion of the results, and the writing of the manuscript.

\paragraph*{Competing interests:}
The authors declare that they have no competing interests.

\paragraph*{Data and materials availability:}
All the data needed to evaluate the conclusions in the paper are present in
the paper and/or the Supplementary Materials.

\end{document}